# Fiber-Based Focal Plane Array Beamformer as Air Interface of an Alignment-Tolerant Optical Fi-Wi-Fi Bridge

Florian Honz, and Bernhard Schrenk

AIT Austrian Institute of Technology, 1210 Vienna, Austria. Author e-mail: florian.honz@ait.ac.at

**Abstract** We demonstrate robust light coupling between two single-mode fibers for an out-door FSO link through a focal plane array beamformer with 61 fine-pitched fiber cores as antenna elements. We show that favourable coupling conditions are established for this 10Gb/s Fi-Wi-Fi bridge after rough initial pointing.

**Introduction**
The continuity in Internet traffic explosion and the fact that communication backbones are facing scalability limits [1] renders the efficient use of bandwidth of utmost importance. Optical communication networks can build on wideband multiplexing techniques [2], provided that fiber as the arguably most attractive transmission medium is available. Fiber-scarce environments, on the other hand, have to rely on alternative options such as THz technology [3], which might face severe limitations due to high channel loss, limited transmit power and bandwidth [4]. To this end, free-space optical (FSO) communication serves as a fiber-grade solution to extend the bandwidth continuum, demonstrating capacities of more than 14 Tb/s [5] or 1.1 Tb/s/λ by means of mode multiplexing [6]. However, these high capacities come at the expense of complex FSO beam shaping, acquisition and tracking [7, 8], due to the necessary optical coupling between fiber and air. This renders FSO links as an economically unattractive option. In order to simplify such FSO systems, MIMO schemes [9, 10], PIC-based beamformers [11-13] and large-core optics [14] have been investigated to address the robustness to atmospheric effects and long-term stability through enhancing the tolerance to alignment errors.

In this work, we experimentally demonstrate a focal plane array (FPA) configuration as an optical beamformer at each end of a point-to-point FSO link. A fiber-based FPA layout with standard single-mode fiber (SMF) is chosen to obtain an air interface between two field-installed fiber trunks. We show that the FPA is capable of obtaining and maintaining optimized coupling conditions after finding "first light" through a coarse manual setup of the optical antennas, leading to error-free operation for an out-door roof-top link operated at 10 Gb/s.

**Focal Plane Array as Air Interface in FSO Link**
In contrast to alternative beamforming methods, FPAs feature a simple control and calibration [15]. Figure 1(a) presents our fiber-based FPA used as the air interface of a point-to-point FSO link between two ITU-T G.652B-compatible SMFs. The focal plane features 61 individual SMF cores of a photonic lantern used as antenna elements, which are spaced by a fixed 37-µm pitch ($\Pi$ in Fig. 1(a)) in a hexagonal arrangement that is centered to an optical lens with focal length $f$. In this way, an offset between the focal center of the lens and the actually illuminated core determines the beam steering angle of the FPA, which over all $N$ antenna elements in a row of the hexagonal arrangement leads to a total field-of-illumination according to $FoI = N \cdot \Pi / f$. The fill factor for all cores over the focal plane is 5.4% or -12.7 dB.

The spatial distribution of the cores over the focal plane enables flexible switching of the optical pencil-beam angle via a change of the used transmitter (and receiver) cores $\zeta_{TX}$ (and $\xi_{RX}$). Thereby, the field-of-illumination (and field-of-view) is enhanced compared to a single fiber core, eventually enabling SMF coupling

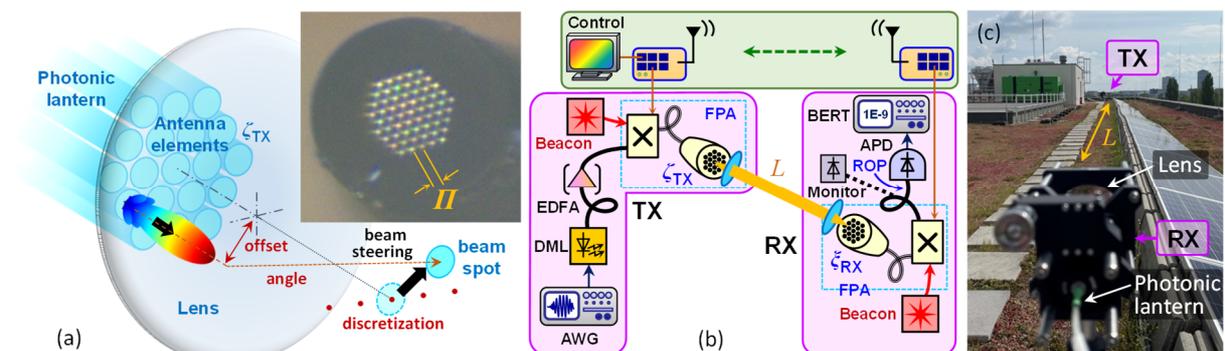

**Fig. 1:** (a) FPA beamformer and photograph of the antenna elements. (b) Setup and (c) installation for out-door FSO link.

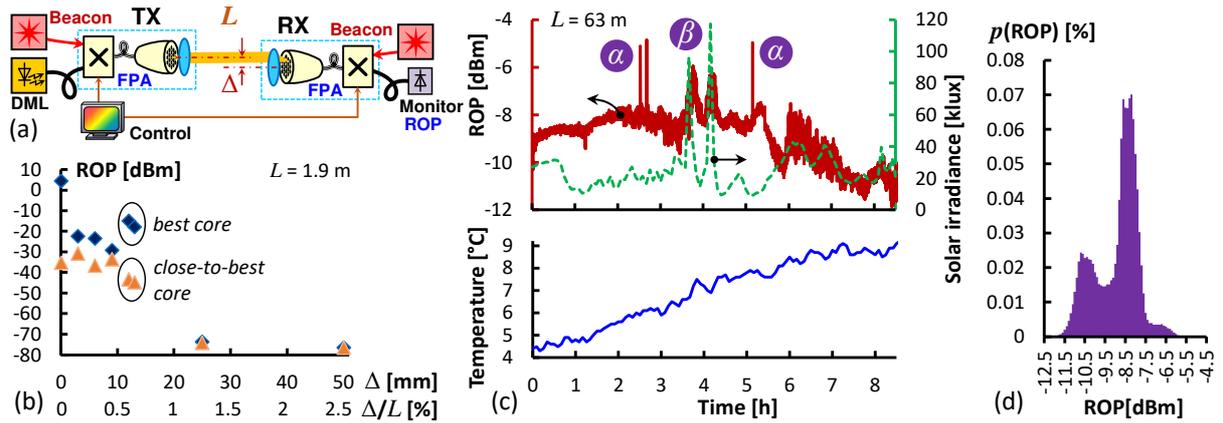

**Fig. 2:** (a) Characterization setup. (b) Alignment tolerance to lateral offset Δ. (c) Long-term link stability and (d) ROP histogram.

and data transmission even for unfavourable initial alignment conditions.

**Discrete Beam Steering and Tolerable Offset**

For a first FPA characterisation we fed the transmitter with a 1550-nm laser having an output power of 10 dBm (Fig. 2(a)). The transmitter steers the FPA beam through a 1×32 MEMS switch, which is connected to the inner 32 cores of the hexagonal antenna element array. The photonic lantern following the optical switch features a low average loss of 0.4 dB for its fiber cores. The output light at the focal plane is then collimated by a static lens with a diameter of 6 mm, having a focal length of $f$ = 10 mm. This yields a pencil beam that is then received by a similar FPA arrangement for which a 64×1 MEMS switch is selecting the receiving antenna element. We monitored the received optical power (ROP) to analyse the coupling efficiency after initial pointing of the FPAs, which is facilitated through red beacon lasers. The distance $L$ between the FPAs for this in-door setup was 1.9 m.

After fine alignment through the two actuated FPAs, we achieved a very low coupling loss of 1.6 dB for the optimal face-to-face coupling between the central transmitter and receiver cores ($31_{TX} \rightarrow 31_{RX}$), with additional 1.6 dB of loss incurring per MEMS switch. Due to the short focal length of the lens, the discretization when steering the collimated pencil beam is ~7 mm at the given link distance of $L$ = 1.9 m. This explains the drop of 26.8 dB in ROP in Fig. 2(b) for a lateral receiver offset of Δ = 3 mm from the initial face-to-face position, since with the given beam diameter of 2.8 mm there is no ideal antenna element available. For an increased offset the ROP decreases slightly, until a local maximum of -15 dBm is reached at an offset of Δ = 12 mm or Δ/$L$ = 0.63% of the FSO reach. The accompanying decrease in ROP for the second-best core confirms the improved coupling conditions at this Δ value. The coupling drops significantly for larger lateral offset values since the number of cores is not large enough to support a wider field-of-view for the given optical FPA parameters. Furthermore, this initial characterization emphasizes the criticality of mismatch between the small beam diameter and the discretization in beam steering for this in-door link. For a given number of cores (yielding $N$), it is therefore required to optimize the lens parameters and thus the beam diameter according to the fixed antenna element pitch $Π$, defining together with the focal length $f$ the discretization of the FPA beam steering and the maximum offset in misalignment that can be recovered. Given the layout of a short-range FSO application for an out-door fiber-wireless-fiber (Fi-Wi-Fi) bridge, the maximum offset that needs to be accounted for after coarse initial pointing of the optical antennas is clearly less critical than the discretization in beam steering.

**Out-door Fi-Wi-Fi Bridge**

The roof-top FSO link investigated for the out-door Fi-Wi-Fi bridge is presented in Fig. 1(b). Given the reach of $L$ = 63 m for the link, the FPA optics were changed to 2"-lenses with a focal length of $f$ = 100 mm and an f-number of f/2, resulting in a beam diameter of ~28 mm and a discrete step size of ~23 mm when steering the beam. This provides a better FPA setup since the individual beam spots impinging upon the receiving FPA do now overlap.

We performed a coarse sequential pointing for the optical antennas towards each other (Fig. 1(c)), using a red beacon laser and a visual alignment at the opposite link end. We then launched an EDFA-boosted laser emission at 20 dBm into the transmitting FSO terminal and performed a beam steering sweep for both FPAs at the FSO transmitter and receiver sites. For this, a local controller at the FSO transmitter is remotely connected to the FSO receiver through a WiFi link. Results for this alignment mechanism are presented in Fig. 3(a) in terms of ROP for all pairs of antenna elements. Two favourable pairs can be identified: $30_{TX} \rightarrow 31_{RX}$ and $38_{TX} \rightarrow 31_{RX}$, with a ROP of -8.7 and -9.5 dBm, respectively. These values correspond to a fiber-to-fiber FSO link loss of 28.7 … 29.5 dB. A continuous long-term ROP acquisition was then conducted over a duration of 8 hours on a cold Central European day in April. As Fig. 2(c) proves, stable coupling between SMF ports was maintained without the need for repeated beam steering or antenna element switching. The coupling is not directly impacted by a change in ambient temperature, which showed a swing of 4°C over the entire duration, but on the momentary heating due to the solar influx. With clear-sky conditions during noon ($\beta$) and the late afternoon, the sun heats up the opto-mechanics at the FSO transmitter and receiver, leading to a (positive) change in coupling of up to 2 dB due to thermal expansion. Moreover, we attribute the three short bursts of +3 dB in ROP ($\alpha$) to sudden wind gusts. The histogram of the ROP in Fig. 2(d) shows a probability peak at -8.2 dBm for the ROP distribution, without deep fading. The ROP finds itself in an interval of 3.2 dB during 90% of the time, between -10.9 and -7.7 dBm.

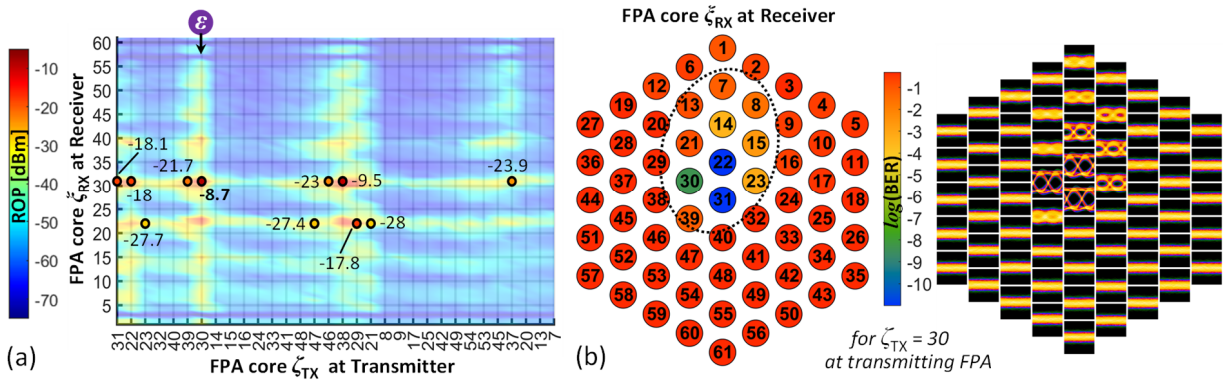

**Fig. 3:** (a) ROP for all 32×64 antenna element pairs $\zeta \to \xi$. (b) Spatial BER distribution and 10-Gb/s eyes over receiving FPA.

**Fi-Wi-Fi Out-door Data Transmission**

The FSO bridge was further evaluated for 10-Gb/s on-off keyed data transmission using a directly modulated laser (DML). The data signal was fed from our laboratory through 360 m of SMF28 trunk fiber to the transmitting FSO terminal at the roof-top, where it is boosted by an EDFA. We employed the same FPA optics as used in the earlier investigation of the link stability. After pencil-beam transmission over the 63-m free-space link and coupling to the receiving FSO terminal, the signal was relayed through another 200 m of trunk fiber back to the lab, where it is received by an avalanche photodetector (APD) with a sensitivity of -28.7 dBm at a BER of $10^{-10}$. We then evaluated the bit error ratio (BER) through a BER tester.

According to the optimal coupling among the FSO terminals, we chose a fixed core $\zeta_{TX} = 30_{TX}$ for the transmitting FPA ($\varepsilon$ in Fig. 3(a)) while varying the core $\xi_{RX}$ of the receiving FPA to investigate a change in the field-of-view. Given the large beam spot, there are multiple $\zeta \to \xi$ pairs that yield a ROP larger than the APD sensitivity. Some of the 18 combinations with >6 dB margin before a relay over the receiver-side trunk fiber are exemplarily highlighted in Fig. 3(a). Specifically, we achieve a BER below $10^{-11}$ for a ROP of -14 and -23.8 dBm at $\xi_{RX} = 31_{RX}$ and $22_{RX}$, respectively (Fig. 4). The elliptic shape in the ROP distribution and open eye diagrams in Fig. 3(b), involving more than five antenna elements, indicate a slight angular pointing error between the transmitting and receiving FPAs. This can be expected since the initial pointing was purely based on a visual alignment of visible laser beacons. It nonetheless proves that a margin of at least 9.8 dB in ROP can be accomplished towards the sensitivity level of the APD receiver. When removing the EDFA booster before the FSO transmitter, we can still achieve transmission at a BER below $10^{-11}$ for $\xi_{RX} = 31_{RX}$ at a ROP just above -28 dBm.

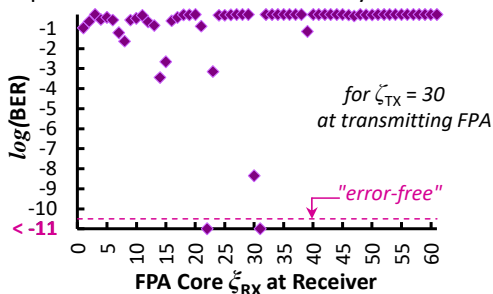

**Fig. 4:** BER for all antenna elements of the receiving FPA.

**Conclusions**

We have demonstrated a simple FPA beamformer as air interface that accomplishes SMF-to-SMF coupling after initial pointing. We have proven the stability of an FPA-enhanced out-door FSO link and achieved error-free 10 Gb/s data transmission over a free-space Fi-Wi-Fi bridge with large 10-dB power margin.

*Acknowledgements*

*This research has received funding from the Smart Networks and Services Joint Undertaking (SNS JU) under the European Union's Horizon-Europe research and innovation programme under Grant Agreement No. 101139182.*